\documentclass[aps,twocolumn,prd,showpacs,showkeys,preprintnumbers,superscriptaddress,nobibnotes,floatfix,longbibliography,notitlepage,nofootinbib]{revtex4-1}

\pdfoutput=1

\usepackage{physics}
\usepackage{enumitem}
\usepackage{mathtools}
\usepackage{amsmath}
\usepackage{amsfonts}
\usepackage{amssymb}
\usepackage{slashed}
\usepackage{mathrsfs}
\usepackage{graphicx}
\usepackage{xcolor}
\usepackage{xspace}
\usepackage{placeins}
\usepackage{booktabs}
\usepackage{float}

\newcommand\colvec[3][]{\begin{pmatrix}\ifx\relax#1\relax\else#1\\\fi#2\\#3\end{pmatrix}}
\definecolor{darkmagenta}{rgb}{0.55, 0.0, 0.55}
\newcommand{\beq}{\begin{equation}}
\newcommand{\beqn}{\begin{eqnarray}}
\newcommand{\eeq}{\end{equation}}
\newcommand{\eeqn}{\end{eqnarray}}

\newcommand{\dk}[2][4]{\frac{\dd[#1]{#2}}{(2\pi)^{#1}}}

\newcommand{\Lagrangian}{\mathcal{L}}

\usepackage{array}
\usepackage{bigints}
\usepackage{tabularx}

\usepackage{csquotes}

\usepackage[colorlinks]{hyperref}
\hypersetup{
    colorlinks,
    linkcolor={red!50!black},
    citecolor={blue!50!black},
    urlcolor={blue!80!black}
}
\usepackage{multirow}
\usepackage{upgreek}
\usepackage[capitalise]{cleveref}
\usepackage{soul}

\newcommand{\sign}{\mathrm{sgn}}

\begin{document}
\include{jdefs}

\title{Photon self-energy at all temperatures and densities in all of phase space}

\author{Hugo Sch\'erer}
\affiliation{Department of Physics \& Trottier Space Institute, McGill University, Montr\'{e}al, QC H3A 2T8, Canada}
\author{Katelin Schutz}
\affiliation{Department of Physics \& Trottier Space Institute, McGill University, Montr\'{e}al, QC H3A 2T8, Canada}

\begin{abstract}
\noindent
In an isotropic background comprised of free charges, the transverse and longitudinal modes of the photon acquire large corrections to their dispersion relations, described by the in-medium photon self-energy. Previous work has developed simple approximations that describe the propagation of on-shell photons in plasmas of varying temperatures and densities. However, off-shell excitations can also receive large medium-induced corrections, and the on-shell approximations have often been used in an effort to capture these effects. In this work we show that the off-shell self-energy can be qualitatively very different than the on-shell case. We develop analytic approximations that are accurate everywhere in phase space, especially in classical and degenerate plasmas. From these, we recover the on-shell expressions in the appropriate limit. Our expressions also reproduce the well-known Lindhard response function from solid-state physics for the longitudinal mode.

\end{abstract}
\maketitle

\section{Introduction}

The dispersion relations of photons in a plasma are qualitatively different from their dispersion relations in vacuum. The photon is massless in vacuum, obeying the gapless dispersion relation $\omega = k$. However, in-medium effects modify this dispersion relation, which can be described via an index of refraction or a frequency-dependent effective mass \cite{braaten_neutrino_1993, raffelt_stars_1996}. Furthermore, photons in vacuum have two polarization modes, both transverse to the propagation direction. However, in-medium collective effects give rise to a third, longitudinal photon mode \cite{raffelt_stars_1996,jackson_classical_1975},\footnote{In the literature, in-medium photons are sometimes referred to as plasmons. Sometimes plasmons only refer to the new longitudinal mode. Here we refer to all these modes as photons.} which has its own distinct, medium-dependent dispersion relation. All of these in-medium properties lead to unique phenomena, such as the decay of photons to neutrinos~\cite{braaten_neutrino_1992, Yakovlev:2000jp, raffelt_stars_1996, Raffelt:1999gv} or new particles beyond the Standard Model~\cite{Raffelt:1987np,Dreiner:2013tja, Vogel:2013raa,Dvorkin:2019zdi,Fung:2023euv}.

The in-medium photon behavior is encoded in the thermal photon self-energy $\Pi^{\mu\nu}$, which depends on the ambient temperature and density. The self-energy is also known as the polarization tensor in linear response theory, where induced currents are linear in the photon field $A^\mu$ in Fourier space, $J_\text{ind}^\mu = \Pi^{\mu\nu}A_\nu$~\cite{raffelt_stars_1996, kapusta_finite-temperature_2011}. Thus, it appears in an effective Lagrangian as
\begin{equation}
    \Lagrangian \supset {\textstyle\frac{1}{2}} A_\mu \Pi^{\mu\nu} A_\nu,
\end{equation}
which is akin to a mass term for the photon. Indeed, the real part of $\Pi^{\mu\nu}$ acts as an effective in-medium mass-squared, while the imaginary parts encode the production and absorption of in-medium photons via the optical theorem~\cite{raffelt_stars_1996,weldon_simple_1983,das_finite_1997,laine_basics_2016,Le_Bellac_1996}.

The presence of the medium singles out a preferred reference frame where the medium's bulk velocity is zero. In that reference frame, for a homogeneous and isotropic medium, the longitudinal and transverse modes are decoupled, and the self-energy can thus be completely described by two form factors,
\begin{equation}
    \Pi^{\mu\nu} = -\Pi_L P_L^{\mu\nu} - \Pi_T P_T^{\mu\nu}
\end{equation}
where $P_a^{\mu\nu}$ are projection operators and $a=L,T$ indexes the longitudinal and transverse modes, respectively \cite{raffelt_stars_1996,kapusta_finite-temperature_2011}. It is straightforward to evaluate the form factors in the medium rest frame, and extract the longitudinal and transverse self-energies. In Lorenz gauge\footnote{Note that in Coulomb gauge, the transverse self-energy and propagators are the same as in Lorenz gauge, while the longitudinal ones become $\Pi_L^\text{Coul.} = \Pi^{00}$ and $D_L^\text{Coul.} = i(k^2-\Pi_L^\text{Coul.})^{-1}$ in the medium rest frame~\cite{braaten_neutrino_1993}.}  \cite{braaten_neutrino_1993,raffelt_stars_1996},
\begin{equation}
    \Pi_L = \frac{\omega^2-k^2}{k^2} \Pi^{00}, \qquad
    \Pi_T = \frac{1}{2}\qty(\delta_{ij} - \frac{k_i k_j}{k^2})\Pi^{ij}, \label{eq:PL-PiProjection}
\end{equation}
where the factor of $1/2$ in $\Pi_T$ comes from the fact that we are averaging over two degenerate transverse modes. The propagator can then be decomposed similarly~\cite{kapusta_finite-temperature_2011},
\begin{equation}
    D^{\mu\nu} = D_L P_L^{\mu\nu} + D_T P_T^{\mu\nu}, ~~~~ ~~~D_a = \frac{i}{\omega^2-k^2-\Pi_a}.~~~~
\end{equation}

In general, the photon self-energy at one-loop can be expressed as an integral that can only be computed numerically. However, in 1993 Braaten and Segel computed analytic approximations for the on-shell self-energy of longitudinal and transverse photons \cite{braaten_neutrino_1993}. Remarkably, these approximations are valid for any photon energy and for any ambient temperature and electron density.

Crucially, the approximations of Ref.~\cite{braaten_neutrino_1993} rely on the assumption that the photons are \textit{on shell}, i.e. satisfying their dispersion relations. However, these analytic approximations have frequently been used for computations involving off-shell photon propagators. This often-overlooked fact was recently highlighted in an erratum of Ref.~\cite{raffelt_stars_1996} (see Appendix E of that work online \cite{raffelt_stars_2023}). It is therefore timely to study the behavior of the self-energy in the off-shell regime to determine the extent to which it differs from the on-shell approximations.

In this paper, we derive analytic expressions akin to those of Ref.~\cite{braaten_neutrino_1993} that are valid in all parts of phase space. These approximations can be used to significantly speed up phase space integrals involving off-shell photons in a medium. Furthermore, approximations for the self-energy can provide an analytic handle on the presence of resonances under different environmental conditions. This can be especially useful when considering searches for particles beyond the Standard Model that rely on the resonant level-crossing with photons that can occur in a medium (see e.g. Refs.~\cite{Raffelt:1987im,Fischbach:1994ir,Hochmuth:2007hk,Arias:2012az,An:2013yfc,Redondo:2013lna,Hardy:2016kme,Chang:2016ntp,McDermott:2019lch,Mirizzi:2009iz,Mirizzi:2009nq,Caputo:2020bdy,Li:2023vpv,Garcia:2020qrp,Lawson:2019brd,Gelmini:2020kcu, Brahma:2023zcw}).

The rest of the paper is organized as follows. In Section~\ref{sec:BSA1} we review the computation of the one-loop photon self-energy in a homogeneous, isotropic plasma in thermal equilibrium using the imaginary-time formalism. We then introduce the ``sharp-peak'' approximation used in Ref.~\cite{braaten_neutrino_1993} and develop the framework for its application to off-shell photons in Section~\ref{sec:approx}. We present our approximations for the real and imaginary parts of the self-energy in Sections~\ref{sec:real} and~\ref{sec:imaginary}, respectively, including simplifications that can be made in the classical, degenerate, and relativistic limits. Discussion and concluding remarks follow in Section~\ref{sec:conclusion}.

\section{One-loop self energy integral}
\label{sec:BSA1}
For a system in equilibrium with temperature $T \equiv 1/\beta$, the density matrix can be expressed as $\rho(\beta) = e^{-\mathcal{H} \beta}$ where $\mathcal{H}$ is the effective Hamiltonian. This is mathematically equivalent to unitary time evolution with a time-independent Hamiltonian $U(t_2, t_1) = e^{-i (t_2-t_1) \mathcal{H}}$ if we interpret $\beta$ as an imaginary time,  $\rho(\beta) = U(-i\beta, 0)$. We can then apply insights from regular quantum field theory but restricting the time domain to a finite interval on the imaginary axis. This interval has periodic boundary conditions for bosonic degrees of freedom since the system is in equilibrium, so the initial and final field configurations are the same. For fermions, equilibrium implies anti-periodic boundary conditions since these states anti-commute. Therefore, when Fourier transforming one obtains a discrete spectrum of imaginary Matsubara frequencies, which take the form $\omega_n = 2  \pi n/\beta$ for bosons and $\omega_n = (2n+1) \pi/\beta$ for fermions with $n \in \mathbb{Z}$, at zero chemical potential. A non-zero chemical potential $\mu$ shifts the energy $p_0 = i\omega_n$ to $p_0 = i\omega_n + \mu$. Phase space integrals over the energy then become a discrete sum over these modes. In effect, Feynman rules are the same as in zero-temperature quantum field theory, except that we substitute loop integrals with
\begin{equation}
\int \frac{d^4 p}{(2\pi)^4} M(p_0) \to \frac{i}{\beta} \sum_{n=-\infty}^{\infty} \int \frac{d^3 p}{(2\pi)^3} M(p_0 = i\omega_n),
\end{equation}
for some arbitrary integrand $M(p)$~\cite{das_finite_1997}.

This formalism is useful to compute thermal two-point functions among other things. To extract physical meaning, one must proceed to analytic continuation from discrete imaginary energies to real energies, which can be done in multiple ways. Here, we analytically continue to $\omega + i \epsilon$ in order to obtain the retarded self-energy.

\begin{figure}[!t]
\centering
\includegraphics[width=0.32\textwidth]{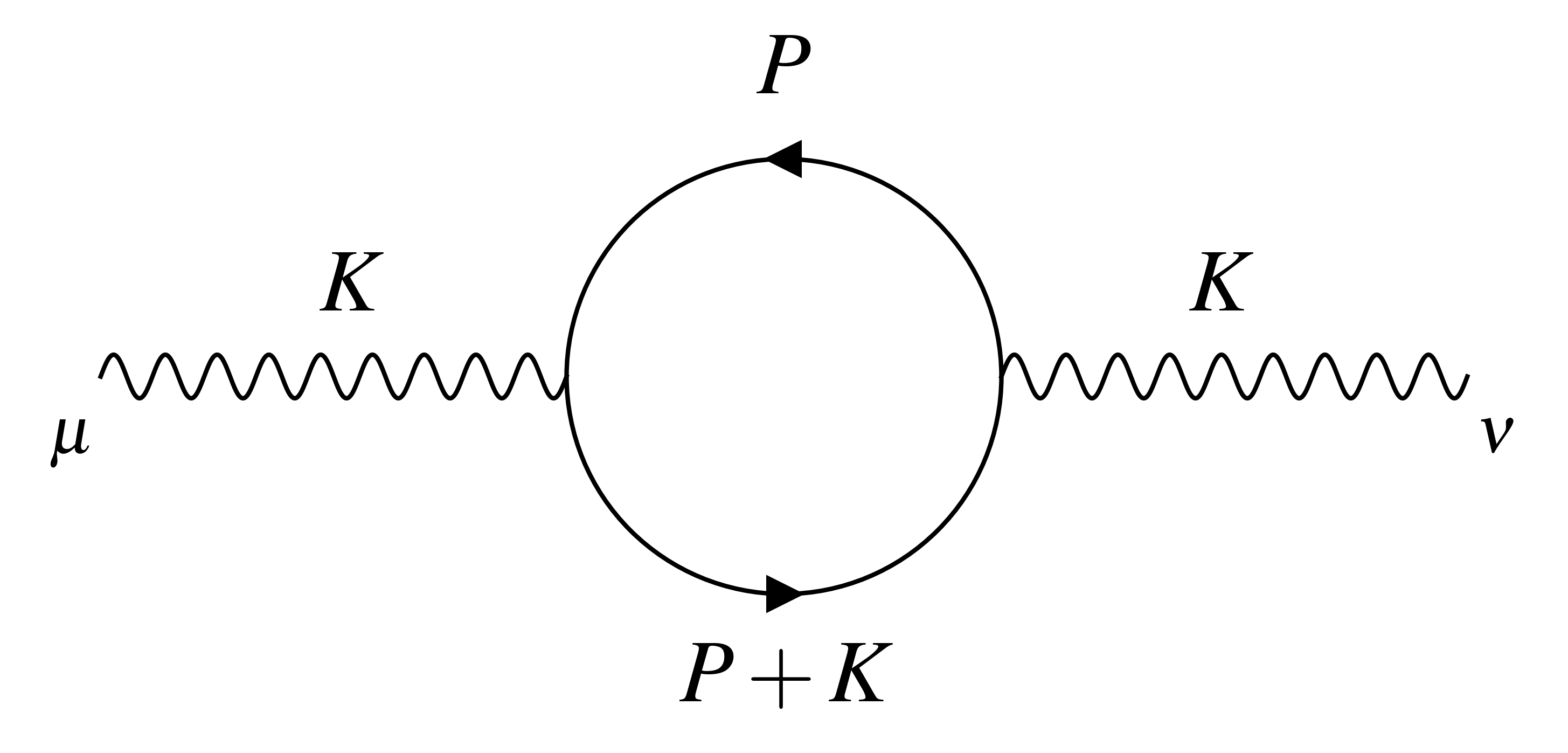}
\includegraphics[width=0.48\textwidth]{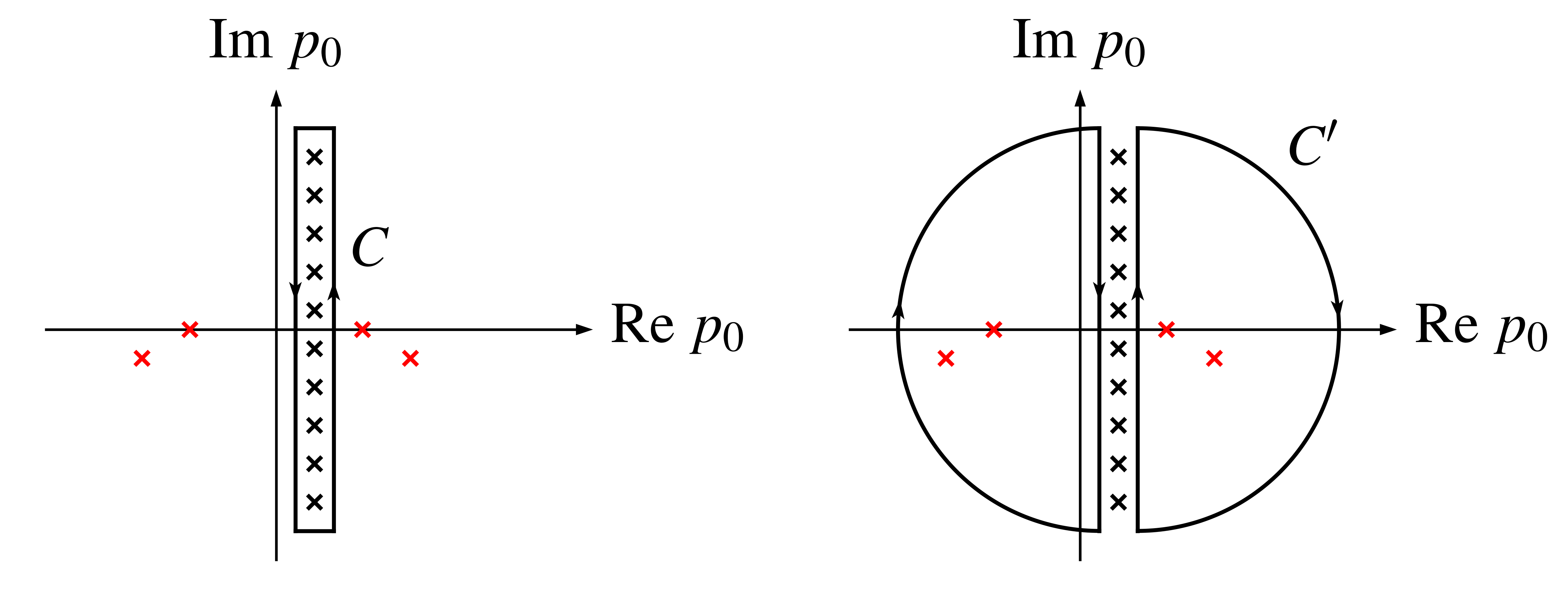}
    \caption{Schematic representations of the one-loop self-energy diagram as well as the contours used to perform the sum over Matsubara modes.}
    \label{fig:schematic}
\end{figure}

The retarded photon self-energy at one-loop order can be readily obtained using the imaginary-time Feynman rules to compute the diagram shown in Figure~\ref{fig:schematic}. For a photon with four-momentum $K^\mu = (k_0, \vec{k})$,
\begin{equation}
    \Pi^{\mu\nu}= 4\pi\alpha T\sum_n \int \frac{d^3 p}{(2\pi)^3} \frac{\Tr[(\slashed{P}+m) \gamma^\mu (\slashed{P}+\slashed{K}+m) \gamma^\nu]}{(P^2-m^2)((P+K)^2-m^2)}
\end{equation}
where $m$ is the mass of the fermion in the loop, $P^\mu = (p_0,\vb{p})$ with $p_0 = (2n+1)  \pi i/\beta + \mu$, and similarly $k_0 = 2 \ell \pi i/\beta$ with $\ell \in \mathbb{Z}$. The infinite sum can be evaluated by realizing that the fermionic Matsubara frequencies are proportional to the infinite number of simple poles of the tanh function. Thus, one can transform this sum into an infinitesimally narrow contour $C$ extending to $\pm \infty$ in the imaginary direction, as shown in Figure~\ref{fig:schematic}. The infinitesimal sides do not contribute, so this is equivalent to an integral over a contour $C'$ with semi-circles closed at infinity, also shown in Figure~\ref{fig:schematic}. This contour $C'$ now only encloses the poles from the fermion propagator from the loop, transforming the infinite sum into a sum of only four terms. Mathematically, for a function $\mathcal{A}(p_0)$ that vanishes at infinity and that has no poles where $\Re(p_0)=\mu$,
\begin{equation}
\begin{aligned}[c]
    T\sum_n \mathcal{A}&(p_0 = (2n+1)  \pi i/\beta + \mu)\\
    &= \frac{1}{2\pi i} \oint_C dp_0 \ \mathcal{A}(p_0) \frac{1}{2} \tanh \bigg(\frac{p_0-\mu}{2T}\bigg)\\
    &= \frac{1}{2\pi i} \oint_{C'} dp_0 \ \mathcal{A}(p_0) \frac{1}{2} \tanh \bigg(\frac{p_0-\mu}{2T}\bigg).
\end{aligned}
\end{equation}
The poles for the $C'$ contour in this case are situated at $p_0$ equal to $E_p$, $-E_p$, $E_{p+k}-k_0$ and $-E_{p+k}-k_0$, where $E_q = \sqrt{\textbf{q}^2 + m^2} > 0$. Using the identity $\tanh(z) = 1 - {2}/({e^{2z} + 1})$, along with other properties of the tanh function, we obtain thermal phase space distributions for the fermions in the loop,
\begin{widetext}
\begin{equation}
\begin{aligned}
    \Pi^{\mu\nu}
    = 16 &\pi \alpha \int \dk[3]{p} \frac{\qty(-1 +f_p+ \bar{f}_p)}{2E_p} \Bigg[
    \frac{\Tr[(\slashed{P}+m) \gamma^\mu (\slashed{P}+\slashed{K}+m) \gamma^\nu]}{4((P+K)^2-m^2)}+\frac{\Tr[(\slashed{P}-\slashed{K}+m) \gamma^\mu (\slashed{P}+m) \gamma^\nu]}{4((P-K)^2-m^2)}\Bigg],
    \label{eq:minusone}
\end{aligned}
\end{equation}
\end{widetext}
where $p_0 = E_p$ now due to shifts in the integration variables for some of the terms, and where $f$ is the Fermi-Dirac distribution,
\begin{equation}
    f_p  =  \frac{1}{e^{(E_p-\mu)/T} + 1},\quad
    \bar{f}_p  = \frac{1}{e^{(E_p+\mu)/T} + 1}.
\end{equation}
The $-1$ term in parentheses diverges for the real part of the self-energy, and should be dealt with through regular renormalization techniques from zero-temperature field theory. The $-1$ term does not diverge for the imaginary part, and should be included in the finite-temperature calculation. The other two terms in parentheses involving the phase space are thermal contributions to both the real and imaginary parts of the self-energy. Evaluating the Dirac traces and proceeding to analytic continuation to evaluate it just above the real line, at $k_0 = \omega + i\epsilon$ (instead of at quantized values of $k_0$ only), yields the often-cited result,
\begin{widetext}
\begin{equation}
    \Pi^{\mu\nu}(\omega, \vb{k})
    = 16 \pi \alpha \int \dk[3]{p} \frac{1}{2E_p}\qty[-1+f_p + \bar{f}_p]
    \frac{(P \cdot K)(K^\mu P^\nu + K^\nu P^\mu)-(K^2)P^\mu P^\nu - (P \cdot K)^2\eta^{\mu \nu}}{(P \cdot K)^2 - \frac{1}{4}(K^2)^2}. \label{BSA1}
\end{equation}
\end{widetext}
where $P^\mu=(E_p, \vb{p})$ and $K^\mu=(\omega + i\epsilon, \vb{k})$, with $\omega$ real and $\epsilon \to 0^+$. This analytic continuation amounts to the retarded self-energy. It is exact at one-loop, and usually constitutes the starting point on photon self-energy discussions across the literature, with the $-1$ term and the $+i\epsilon$ often omitted when discussing the real part only. In this form (and dropping the $-1$), the real part of the thermal self-energy can be readily interpreted as the forward scattering of photons with actual electrons and positrons from the plasma, appropriately weighed with their phase space distributions. The interpretation of the imaginary part is discussed in Section~\ref{sec:imaginary}.

\section{Sharp-Peak Approximation}
\label{sec:approx}
\subsection{Phase Space}
In general, the self-energy is a function of $\omega$ and $k$. However, in some cases it is useful to re-parameterize the phase space in order to make contact with Ref.~\cite{braaten_neutrino_1993}. To this end, we  define
\begin{equation}
    n = \frac{k}{\omega}, \qquad
    \xi = \frac{\sqrt{\abs{\omega^2 - k^2}}}{2m}.
\end{equation}
The first parameter, $n$, is the index of refraction, and it allows us to easily delineate between timelike $(0<n<1)$ and spacelike $(n>1)$ photons. The parameter $\xi$ represents the ratio of the photon gap mass to twice the mass of the lightest charge carriers (usually electrons and positrons are the most relevant). We can invert these relations if necessary for $n \neq 1$, while $n=1$ simply corresponds to the gapless vacuum case. In certain limits, it will be convenient to express the final expressions case by case for three regimes,
\begin{itemize}
    \item timelike light photons $(0<n<1, 0 < \xi < 1)$,
    \item timelike heavy photons $(0<n<1, \xi > 1)$,
    \item and spacelike photons $(n>1)$.
\end{itemize}
Furthermore, note that $k = \abs{\vb{k}}$ is positive, while $\omega$ can be positive or negative. Since the real part of self-energy is even in $\omega$ and the imaginary part is odd~\cite{weldon_simple_1983}, we will assume $\omega > 0$ (equivalently $n>0$) and note that the extension to other parts of phase space is straightforward.

\subsection{Transforming the integrand}
The one-loop integral of Eq.~\eqref{BSA1} cannot be evaluated analytically given the form of the phase space distributions. However, Ref.~\cite{braaten_neutrino_1993} effectively employed the fact that the derivative of the Fermi-Dirac distribution is sharply peaked around a specific velocity. Therefore, using integration by parts, one can derive analytic approximations for the one-loop self-energy. In this Section, we review their approach and its application to on-shell photons, which we then use to derive our general expressions which are also valid off shell.

The self-energy integrand is independent of the azimuthal angle and depends only on $\cos \theta = (\vec k \cdot \vec p\,)/k p$. We introduce the function $G_a$, with $a = T,L$, so that we can express the self-energy as
\begin{equation}
    \Pi_a(K) = \frac{4 \alpha}{\pi} \int_0^\infty \dd{p} \frac{p^2}{2 E_p} (f_p + \bar{f}_p)  \int_{-1}^1 \dd{(\cos \theta)} G_a.
\end{equation}
Projecting Eq.~\eqref{BSA1} onto the transverse and longitudinal directions, we obtain
\begin{subequations}
\begin{align}
    G_L &= \frac{K^2}{k^2} \frac{2 E_p \omega(P \cdot K) - K^2 E_p^2 - (P \cdot K)^2}{(P \cdot K)^2 - (K^2)^2/4},\\
    G_T &= \frac{- K^2 p^2 \qty(1 - \cos^2\theta)/2 + (P \cdot K)^2}{(P \cdot K)^2 - (K^2)^2/4}.
\end{align}
\label{eq:Gs}
\end{subequations}
We define \begin{equation}
     F_a(v) = \frac{v^2}{2(1-v^2)^2} \int_{-1}^{1} \dd{(\cos\theta)} G_a,
\end{equation}
so that after re-expressing the self-energy as an integral over $v = p/E_p$, we obtain
\begin{equation}
    \Pi_a = m^2 \frac{4 \alpha}{\pi} \int_0^1 \dd{v} (f_p + \bar{f}_p) F_a(v).
\end{equation}
Further, we can make contact with the plasma frequency
\begin{align}
    \omega_p^2 &\equiv m^2 \frac{4 \alpha}{\pi} \int_0^1 \dd{v} \frac{v^2}{(1-v^2)^2} \qty(1-\frac{1}{3}v^2) (f_p + \bar{f}_p)\\
    &=  - m^2 \frac{4 \alpha}{3\pi} \int_0^1 \dd{v} \frac{v^3}{1-v^2} \dv{v}(f_p + \bar{f}_p)
    = \int_0^1 \dd{v} \dv{\omega_p^2}{v}\quad  \quad \qquad \nonumber
\end{align}
by defining
\begin{equation}
    J_a(v) = \frac{3(1-v^2)}{v^3} \int_0^v \dd{v'} F_a(v'),
\end{equation}
so that integrating the self-energy by parts yields
\begin{equation}
    \Pi_a = \int_0^1 \dd{v} J_a(v) \dv{\omega_p^2}{v}.
\label{eq:sharp}
\end{equation}
Note that in some instances of integration by parts (particularly when evaluating the real part of the self-energy), the boundary terms vanish either manifestly or because of the limiting behavior of the phase space distributions.

Since $f(z) \sim \tanh(z)$, the derivative of the phase space should be sharply peaked around some velocity $v_*$, and therefore $\dv*{\omega_p^2}{v}$ should be sharply peaked as well. If this peak is sharp and narrow enough, and if $J_a(v)$ is slow-varying enough around this peak, then the integral only has support at $v= v_*$. We can thus approximate $J_a(v) \simeq J_a(v_*)$ in the integrand and take it out of the integral, leaving
\begin{equation}
    \Pi_a \simeq \omega_p^2 J_a(v_*).
\end{equation}
The velocity $v_*$ is given by $v_* = \omega_1/\omega_p$ with
\begin{equation} \label{eq:BS-omega1sq}
    \omega_1^2 = \frac{4\alpha}{\pi} \int_0^\infty \dd{p} \frac{p^2}{E_p} \qty(\frac{5}{3}v^2 - v^4) \qty(f_p + \bar{f}_p).
\end{equation}
and is the approximate velocity where the integrand is peaked. It is defined such that $0<v_*<1$, and physically represents the typical velocity of an electron or positron in the plasma.

The procedure described above can be straightforwardly applied to the real part of the self-energy. The imaginary part is best computed through cutting rules, similar to the optical theorem in zero-temperature field theory \cite{weldon_simple_1983,das_finite_1997,Le_Bellac_1996}. Both fermions in the loop are thus put on shell, and the integral is restricted by kinematic considerations, as detailed in Section~\ref{sec:imaginary}. Crucially, this means that the boundary terms of Eq.~\eqref{eq:sharp} do not necessarily vanish for the imaginary part of the self-energy. However, the sharp peak approximation still applies whenever $v_*$ is within the kinematically allowed velocities.

\subsection{On-shell case}
For a general point in phase space, we must keep the $(K^2)^2$ term in the denominator of Eq.~\eqref{BSA1}. However, as argued in Ref.~\cite{braaten_neutrino_1993}, we can remove this term for on-shell photons and repeat the procedure outlined above. Removing the $(K^2)^2$ term is a convenient way to remove the unphysical imaginary part of the on-shell self-energy proportional to the decay rate for $\gamma \to e^+ e^-$. This imaginary part appears when both the electrons in the loop go on shell, which corresponds to the kinematics where the denominator goes to zero. In fact, thermal corrections to the electron (and positron) mass always guarantee that the decay $\gamma \to e^+ e^-$ is kinematically forbidden for on-shell particles \cite{kapusta_finite-temperature_2011,braaten_neutrino_1992}, which is not explicitly captured by Eq.~\eqref{BSA1}. Even off shell, the $(K^2)^2$ term can also be neglected in some kinematic regimes. There is a separation of momentum scales for on-shell photons which amounts to $\abs{K^2} \ll \abs{2 P\cdot K}$. These expression for on-shell photons are thus also valid in the soft-momentum limit for off-shell photons.

After dropping this term from Eq.~\eqref{eq:Gs} and integrating over the angle, we obtain
\begin{align}
    F_L^\text{On} &= \frac{v^2 (1-n^2)}{n^2(1-v^2)^2} \qty[ \frac{1}{nv}\log \frac{1+nv}{1-nv} - \frac{1-n^2}{1-n^2v^2} -1 ]\nonumber\\
    F_T^\text{On} &= - \frac{v^2(1-n^2)}{2n^2(1-v^2)^2} \qty[ \frac{1}{nv}\log \frac{1+nv}{1-nv}-\frac{2}{1-n^2} ].~~~~~~~
\end{align}
For on-shell photons, the self-energy thus ends up depending on $k$ and $\omega$ only via the dependence on $n$. Performing the velocity integral and using the sharp-peak approximation, we find
\begin{subequations}
\begin{align}
    \Pi_L^\text{On} &=  \frac{3\omega_p^2}{v_*^2} \qty(\frac{1-n^2}{n^2}) \qty[\frac{1}{2nv_*} \log \qty(\frac{1+n v_*}{1-n v_*}) - 1],\\
    \Pi_T^\text{On} &=  \frac{3\omega_p^2}{2 v_*^2}\qty[\frac{1}{n^2} - \qty(\frac{1-n^2v_*^2}{n^2}) \frac{1}{2nv_*} \log \qty(\frac{1+n v_*}{1-n v_*})].
\end{align}
\end{subequations}
These are precisely the results obtained in Ref.~\cite{braaten_neutrino_1993}. These expressions implicitly depend on the properties of the plasma that affect $\omega_p$ and $v_*$.

For off-shell photons whose energy and momenta are unconstrained, however, $\gamma \to e^+ e^-$  is kinematically allowed if $\xi >1$. Therefore, we must include the $(K^2)^2$ term in the denominator of Eq.~\eqref{BSA1}
in order to obtain an approximation for the self-energy that is valid throughout all of phase space, for any value of $\xi$. Note that the $(K^2)^2$ term in the denominator is proportional to $\xi$, so in the Sections that follow one can recover the on-shell expression by setting $\xi \to 0$.

\begin{widetext}
\section{Real part}
\label{sec:real}
\subsection{General expressions}
Keeping the $(K^2)^2$ in the denominator of Eq.~\eqref{eq:Gs} and performing the integrals using the sharp-peak approximation, we find
\begin{subequations}\label{eq:RePi}
\begin{align}
    &\begin{aligned}
    &\Pi_L = \omega_p^2 \Bigg[ -\frac{2 {K^2}}{k^2 v_*^2}
    +\frac{{K^2}}{4E_*^2v_*^3}\log (\frac{1+v_*}{1-v_*})+\frac{\omega{K^2}\qty(3+(\omega^2-3 k^2)/4E_*^2)}{4 k^3 v_*^3}
    \log \qty|\frac{(\omega+k v_*)^2-(K^2)^2/4E_*^2}{(\omega-k v_*)^2-(K^2)^2/4E_*^2}|\\
    &-\frac{E_*{K^2}(1+3 K^2/4E_*^2)}{2 k^3 v_*^3}
    \log \qty|\frac{\omega^2-(kv_*-K^2/2E_*)^2}{\omega^2-(k v_*+K^2/2E_* )^2}|-\frac{ (1-v_*^2+K^2/2E_*^2) }{2 v_*^3}\sqrt{\abs{\frac{4m^2}{K^2}-1}} C \Bigg]
    \end{aligned}\\
    &\begin{aligned}
    &\Pi_T = \omega_p^2 \Bigg[\frac{k^2+2\omega^2}{2 k^2 v_*^2}
    +\frac{K^2}{4E_*^2v_*^3} \log (\frac{1+v_*}{1-v_*})-\frac{\omega \qty(3(\omega^2-k^2 v_*^2) + (\omega^2+3k^2)K^2/4E_*^2)}{8 k^3 v_*^3}
    \log \qty|\frac{(\omega+k v_*)^2-(K^2)^2/4E_*^2}{(\omega-k v_*)^2-(K^2)^2/4E_*^2}|\\
    &+\frac{E_* \qty(3(\omega^2 - v_*^2k^2) - 2K^2 + 3(\omega^2+k^2)K^2/4E_*^2 )}{4 k^3 v_*^3}
    \log \qty|\frac{\omega^2-(kv_*-K^2/2E_* )^2}{\omega^2-(kv_*+ K^2/2E_*)^2}|-\frac{(1-v_*^2 + K^2/2E_*^2) }{2 v_*^3} \sqrt{\abs{\frac{4m^2}{K^2}-1}}  C\Bigg]
    \end{aligned}
\end{align}
\end{subequations}
where
\begin{equation}
    C =
    \begin{cases}
    \displaystyle
    \tan ^{-1} \qty(\frac{\bigl((K^2)^2/4m^2+k^2\bigr)v_*-\omega k}{\bigl((K^2)^2/4m^2\bigr)\sqrt{4m^2/K^2-1}})
    +\tan ^{-1}\qty(\frac{\bigl((K^2)^2/4m^2+k^2\bigr)v_*+\omega k}{\bigl((K^2)^2/4m^2\bigr)\sqrt{4m^2/K^2-1}}) &\quad {n<1 \text{ and } \xi<1}\\
    \displaystyle
    \frac{1}{2}\log \abs{\frac{\Bigl(v_* \bigl((K^2)^2/4m^2+k^2\bigr)+\bigl((K^2)^2/4m^2\bigr)\sqrt{1-4m^2/K^2}\Bigr)^2-\omega^2k^2}{\Bigl(v_* \bigl((K^2)^2/4m^2+k^2\bigr)-\bigl((K^2)^2/4m^2\bigr)\sqrt{1-4m^2/K^2}\Bigr)^2-\omega^2k^2}} & \quad {n>1 \text{ or } \xi>1}
    \end{cases}
\end{equation}
and $E_* = m/\sqrt{1-v_*^2}$.
\subsection{Limiting Behaviour}
\subsubsection{Kinematics}
Taking the limits $\xi \to 0$ or $n \to 1$ for both the timelike and spacelike expressions, we recover the on-shell self-energy, $\Re \Pi_a(\xi \to 0) = \Pi_a^\text{On}$ for $a=L,T$. The $n \to 0$ limit, corresponding to $k=0$, yields
\begin{equation}
    \Re \Pi({n \to 0}) = \frac{\omega_p^2}{v_*^2}
    + \frac{\omega_p^2(1-v_*^2) \xi ^2}{v_*^3}\log (\frac{1+v_*}{1-v_*})-\frac{\omega_p^2(1-v_*^2) (1+2\xi^2)}{2v_*^3 \xi} \times     \begin{cases}
    \displaystyle 2\sqrt{1-\xi ^2}\tan^{-1}\qty(\frac{v_* \xi}{\sqrt{1-\xi ^2}}) &\text{for } \xi < 1,\\
    \displaystyle \sqrt{\xi ^2-1}\log\qty|\frac{v_* \xi + \sqrt{\xi ^2-1} }{v_* \xi - \sqrt{\xi ^2-1}}| &\text{for } \xi \geq 1.
    \end{cases}
\end{equation}
for both the transverse and longitudinal modes. The $n \to \infty$ limit, corresponding to $\omega=0$, yields
\begin{subequations}
\begin{align}
&\begin{aligned}
    \Re \Pi_L(n \to \infty)= \omega_p^2 \Bigg[&\frac{2}{v_*^2}
    -\frac{(1-v_*^2) \xi^2}{v_*^3}
    \log (\frac{1+v_*}{1-v_*}) -\frac{1-3 (1-v_*^2) \xi ^2}{2 v_*^3 \xi \sqrt{1-v_*^2}}
    \log \abs{\frac{v_*-\sqrt{1-v_*^2} \xi}{v_*+\sqrt{1-v_*^2} \xi}}\\
    &\qquad -\frac{(1-v_*^2) (1-2 \xi ^2) \sqrt{\xi ^2+1}}{2 v_*^3\xi }
    \log \abs{\frac{v_* \sqrt{\xi ^2+1}+\xi}{v_* \sqrt{\xi ^2+1}- \xi }}
    \Bigg]
\end{aligned}\\
&\begin{aligned}
    \Re \Pi_T(n \to \infty)= \omega_p^2 \Bigg[&\frac{1}{2 v_*^2}
    -\frac{(1-v_*^2) \xi ^2 }{v_*^3}
    \log (\frac{1+v_*}{1-v_*})-\frac{2-3v_*^2-3\xi^2(1-v_*^2)}{4 v_*^3 \xi\sqrt{1-v_*^2}}
    \log \abs{\frac{v_*-\sqrt{1-v_*^2} \xi}{v_*+\sqrt{1-v_*^2} \xi}}\\
    &\qquad -\frac{(1-v_*^2) (1-2 \xi ^2) \sqrt{\xi ^2+1} }{2 v_*^3 \xi }
    \log \abs{\frac{v_* \sqrt{\xi ^2+1}+\xi}{v_* \sqrt{\xi ^2+1}- \xi }}
    \Bigg].
\end{aligned}
\end{align}
\end{subequations}
\end{widetext}
\subsubsection{Plasma Properties}\label{sec:PlasmaLimitsReal}
The expressions of Eq.~\eqref{eq:RePi} depend on the ambient plasma properties via the dependence on $\omega_p$ and $v_*$, which can be approximated in different regimes.

In the classical limit,  the plasma is both non-relativistic $m/T \gg 1$ and non-degenerate $(m - \mu)/T \gg 1$. In this case, $\omega_p$ and $v_*$ can be computed to first order~\cite{braaten_neutrino_1993},
\begin{equation}
    \omega_p^2 = \frac{4\pi\alpha n_e}{m} \qty(1-\frac{5T}{2m})\quad \quad
    v_*^2 = \frac{5T}{m} \ll 1
\end{equation}
where $n_e = 2(mT/2\pi)^{3/2} e^{-(m-\mu)/T}$ is the electron density. To leading order in $v_*$ (equivalently, in $T/m$), we find
\begin{widetext}
\begin{subequations}
\begin{align}
   \quad \quad &\Re\Pi_L = \omega_p^2 \frac{4m^2 K^2}{4m^2\omega^2-(K^2)^2}  \Bigg[1 +  \frac{48m^4\omega^2 k^2 + 4m^2(k^2-3\omega^2)(K^2)^2 + 3(K^2)^4}{5(4m^2\omega^2-(K^2)^2)^2}v_*^2\Bigg]\nonumber\\
   \quad \quad & \Re\Pi_T = \omega_p^2 \frac{4m^2\omega^2}{4m^2\omega^2-(K^2)^2}\Bigg[1 +
\frac{16m^4\omega^2k^2 + 4m^2(k^2-3\omega^2)(K^2)^2 + (3+2k^2/\omega^2)(K^2)^4}{5(4m^2\omega^2-(K^2)^2)^2}v_*^2\Bigg].\nonumber
\end{align}
\end{subequations}
\end{widetext}

Meanwhile, in either the high-$T$ or high-$\mu$ limits, \mbox{$v_*\rightarrow 1$} and $\Re\Pi_a \to \Pi_a^\text{On}$, even for off-shell photons. If $\mu=0$, then to leading order in $m/T$ we have
\begin{equation}
    \omega_p^2 = \frac{4\alpha}{3\pi}\qty(\frac{\pi^2}{3}T^2 - m^2)\quad \quad
    1- v_*^2 = \frac{m^2}{4\sqrt{3}T^2}.
\end{equation}
~\\
Alternatively, in the degenerate high-$\mu$ limit with Fermi velocity $v_F$, one finds
\begin{equation}
    \omega_p^2 = \frac{4\alpha}{3\pi}\qty(\mu^2-m^2)v_F\quad \quad
    1-v_*^2 = 1-v_F^2 =\frac{m^2}{\mu^2}
\end{equation}
    in the $T=0$ limit~\cite{braaten_neutrino_1993}. The sharp-peak approximation is particularly good in this regime since the phase space distribution becomes an actual step function. For both of these limits, to leading order in $1-v_*^2$, we find
\begin{widetext}
\begin{subequations} \label{eq:high-v}
\begin{align}
    &\begin{aligned}
    \Re\Pi_L &=\frac{3 \omega_p^2 K^2}{k^2}\qty(\frac{\omega}{2k}\log \abs{\frac{\omega+k}{\omega-k}} - 1)
    + \sign(K^2)\omega_p^2\frac{1-v_*^2}{2}\Bigg[8-\frac{9\omega^2}{k^2} +\qty(\frac{16}{3}
    -\frac{2\omega^2}{k^2}) \frac{K^2}{4m^2}
    - \frac{K^2}{2m^2}\log \frac{1-v_*^2}{4}\\
    & \qquad\qquad +\omega\qty(\frac{9K^2 + (\omega^2-3k^2){K^2}/{2m^2} }{2 k^3})
    \log \abs{\frac{\omega+k}{\omega-k}}
    -\qty(1+\frac{K^2}{2m^2}) \sqrt{\abs{\frac{4m^2}{K^2}-1}} C
    \Bigg]
    \end{aligned}\\
    &\begin{aligned}
    \Re\Pi_T &=
    \frac{3\omega_p^2}{2}\qty(\frac{\omega^2}{k^2} - \frac{\omega K^2}{2k^3}\log\abs{\frac{\omega+k}{\omega-k}})
    + \omega_p^2\frac{1-v_*^2}{2}\Bigg[2+\frac{9\omega^2}{2 k^2}+\qty(\frac{10}{3}+\frac{\omega^2}{k^2})\frac{K^2}{4m^2}
    -\frac{K^2}{2m^2} \log \frac{1-v_*^2}{4}\\
    &\qquad \qquad-\omega\qty(\frac{9K^2+6k^2+(\omega^2 +3k^2) {K^2}/{2m^2}}{4k^3})
    \log\abs{\frac{\omega+k}{\omega-k}}
    -\left(1+\frac{K^2}{2m^2}\right) \sqrt{\abs{1-\frac{4m^2}{K^2}}} C
    \Bigg]
    \end{aligned}
\end{align}
\end{subequations}
where in this limit
\begin{equation}
    C = \begin{cases}
    \displaystyle 2\tan ^{-1}\qty(\frac{1}{\sqrt{{4m^2}/{K^2}-1}}) \qq{if} n < 1 \qq{and} \xi < 1\\
    \displaystyle \log\abs{\frac{1+\sqrt{1-{4m^2}/{K^2}}}{1-\sqrt{1-{4m^2}/{K^2}}}} \qq{if} n > 1 \qq{or} \xi > 1.
    \end{cases}
\end{equation}
\end{widetext}

\begin{figure*}[t]
    \centering
\includegraphics[width=\textwidth]{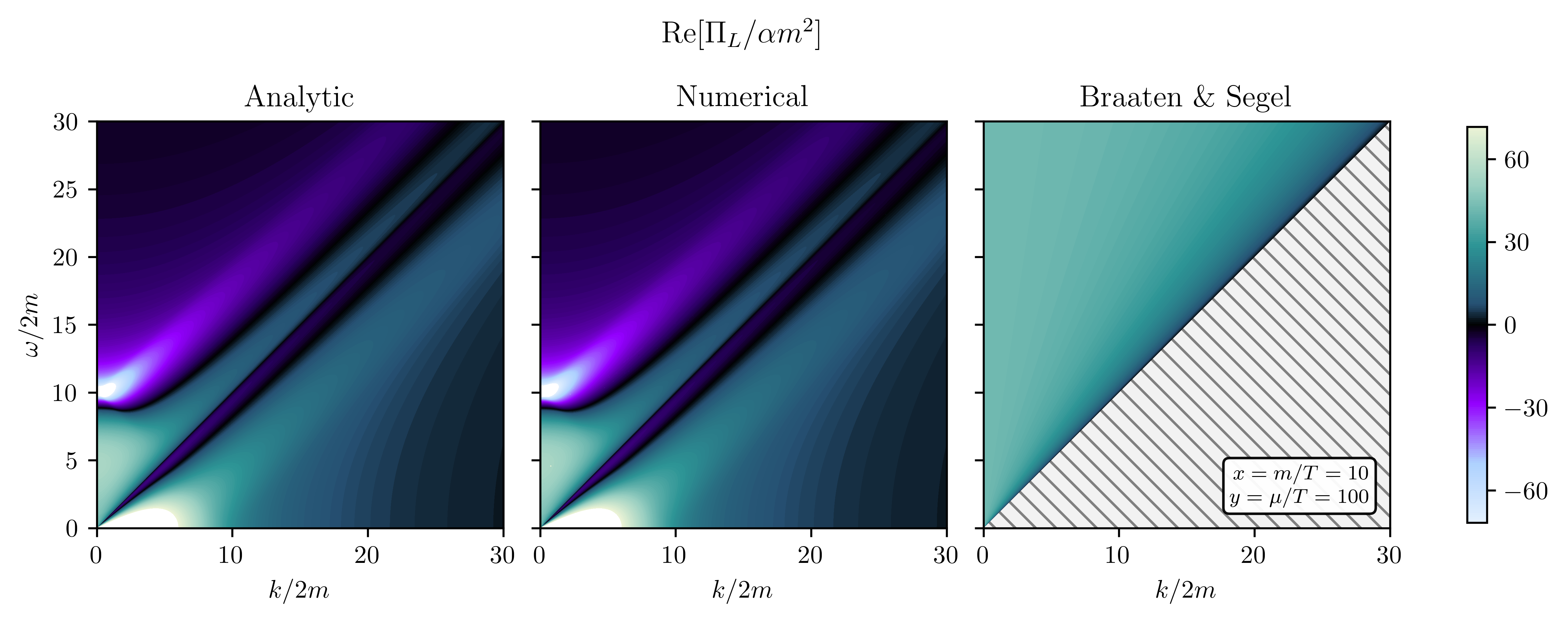}
\includegraphics[width=\textwidth]{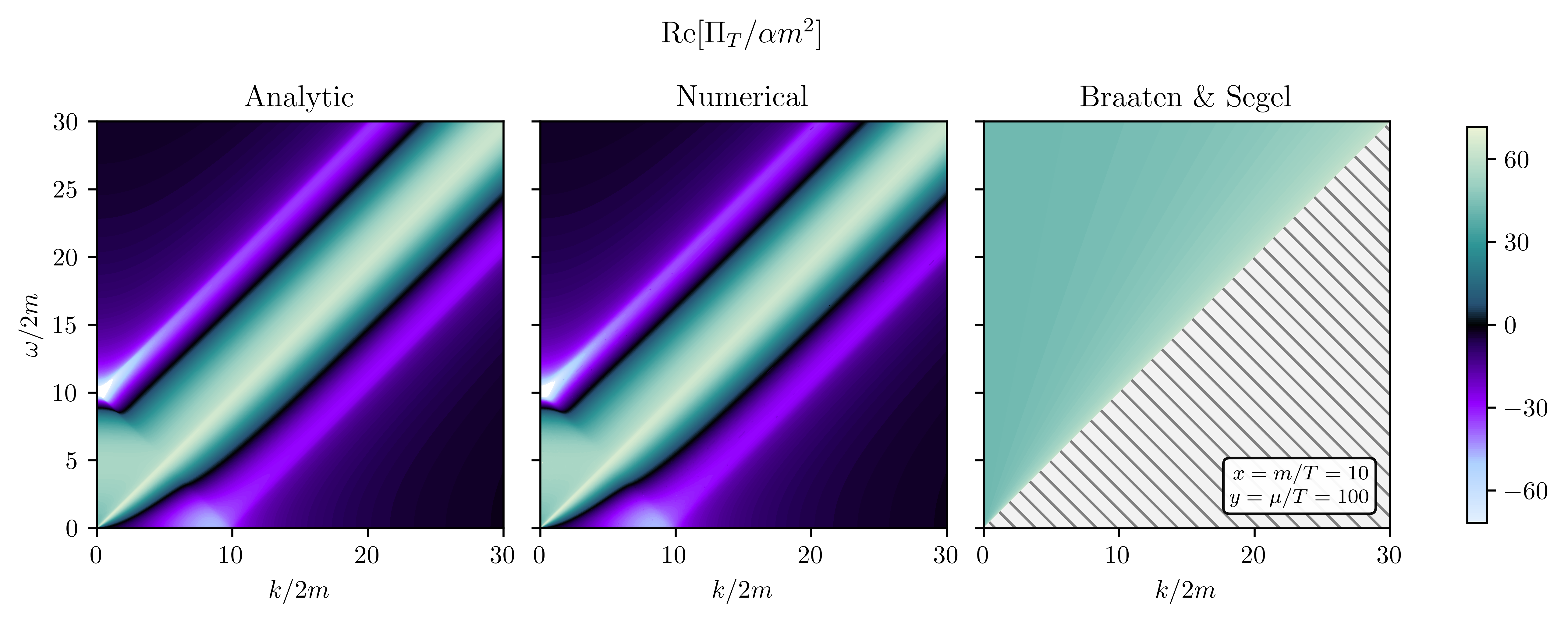}
\vspace{-0.5cm}
    \caption{The real part of the self-energy as a function of $\omega$ and $k$, for longitudinal (top) and transverse (bottom) modes in a plasma where $x=10$ and $y=100$. The self-energy is computed with the analytic approximation $\Pi_a \simeq \omega_p^2 J(v_*)$ (left panels), the numerical integral (middle panels), and with the expression from Ref.~\cite{braaten_neutrino_1993} (right panels).}
    \label{fig:RePi}
\end{figure*}

Finally, in the limit of a degenerate plasma with a small Fermi velocity $v_F \ll 1$, corresponding to $m/T \gg 1$ and $m \lesssim \mu$, we recover Lindhard's function for the longitudinal excitations. We express the dielectric constant $\epsilon_L = 1-\Pi_L/(\omega^2-k^2)$ exclusively in terms of $v_F$, $z = k/2p_F$, $u = \omega/kv_F$, and $\omega_p$, where $p_F = mv_F$ is the Fermi momentum. Taking the leading order contribution in the limit $v_F \ll 1$, we find
\begin{equation}
\begin{aligned}
    \Re[\epsilon_L] &= 1+\frac{3 \omega _p^2}{k^2 v_F^2} \Bigg\{\frac{1}{2}+
    \frac{\qty(1-\qty(z-u)^2)}{8z}  \log \abs{\frac{z-u+1}{z-u-1}}
    \\&+\frac{\qty(1-\qty(z+u)^2) }{8z}
    \log \abs{\frac{z+u+1}{z+u-1}}
    \Bigg\},
\end{aligned}
\end{equation}
which matches Lindhard's function exactly \cite{Lindhard_1954,Dressel_Gruner_2002}. Our expression thus provides a finite-temperature generalization of this well-known result.

\begin{figure*}[t]
    \centering
\includegraphics[width=0.75\textwidth]{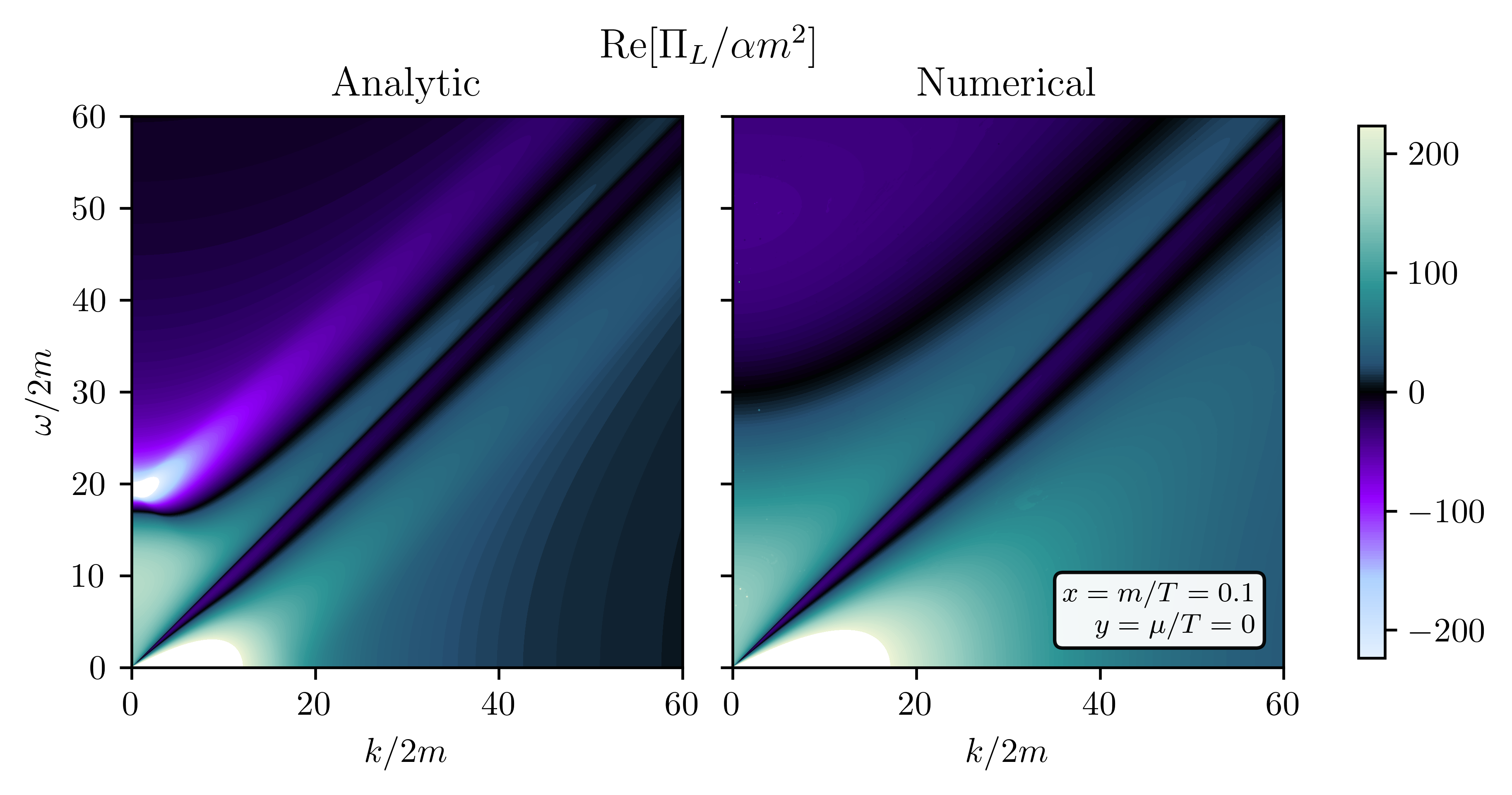}
\vspace{-0.5cm}
    \caption{The real part of the self-energy as a function of $\omega$ and $k$, for longitudinal modes in a plasma where $x=0.1$ and with vanishing chemical potential. The self-energy is computed with the analytic approximation $\Pi_a \simeq \omega_p^2 J(v_*)$ (left panel) and with the numerical integral (right panel).}
    \label{fig:RePiHighT}
\end{figure*}

\subsection{Testing the approximation}
In this Subsection, we compare our analytic approximations to the real part of the one-loop photon self-energy as computed with numerical integration. The integration was orders of magnitude slower to evaluate, and was unstable in part of phase space, further motivating the use of our analytic approximations. The results are shown in Figure~\ref{fig:RePi} for values of $x=m/T=10$ and $y=\mu/T=100$.
These values very roughly correspond to the conditions of a degenerate star, where $x \sim \mathcal{O}(10-100)$ and $y \sim \mathcal{O}(100-1000)$ depending on the star~\cite{raffelt_stars_1996}.

Figure~\ref{fig:RePi} clearly shows that the analytic expressions match the self-energy computed numerically. Quantitatively, we find that in the depicted region of phase space the accuracy is $\mathcal{O}(1\%)$. Furthermore, we confirm that the expression from Ref.~\cite{braaten_neutrino_1993} only approximates the self-energy correctly in the region $\omega \sim k$ where the photon momentum is soft, i.e. $\xi \ll 1$. This is expected since we recover the results of Ref.~\cite{braaten_neutrino_1993} in the limit $\xi \to 0$. Another striking observation is the fact that the self-energy even becomes negative in some parts of phase space, which is qualitatively very different from the on-shell case.

The degree of accuracy of our new analytic expressions depends on the properties of the plasma. For example, in the high-$T$ case $(x \ll 1)$, we find that the analytic expressions are somewhat less accurate in certain localized regions of phase space. In the small-$\xi$ regime we find agreement at the level of $\mathcal{O}(1\%-10\%)$, while in the regime where $\xi \gg 1$ the changes can be $\mathcal{O}(0.1-1)$ especially where $\omega \ll k$ or $k \ll \omega$.
In fact, when $v_* \to 1$ (or equivalently $m/T \to 0$), our high-$T$ expressions Eqs.~\eqref{eq:high-v} appropriately reduce to the hard thermal loop limit, which is known to describe only photons whose energy and momentum are much smaller than the temperature \cite{kapusta_finite-temperature_2011, Le_Bellac_1996}.
However, the differences between the analytic and the numerical expressions are largely driven by differences in the width and position of troughs and peaks in phase space, as can be seen in Figure~\ref{fig:RePiHighT} where $x=0.1$. Therefore, despite the reduced accuracy, the qualitative structure of $\Re \Pi$ is still well-captured by our analytic approximations (including the negative self-energy in some parts of phase space). One could therefore expect that using the analytic approximation to integrate over phase space may still give an accurate overall result, despite local inaccuracies in some kinematic regions. A more quantitative assessment of the accuracy would require using these approximations to calculate the rates of a specific process in a particular environment, which we leave for future work. Regardless, the approximations should drastically reduce the computational cost of performing phase space integrals.

\section{Imaginary part}
\label{sec:imaginary}
\subsection{General expressions}
The imaginary part of the self-energy can be expressed in terms of cutting rules \cite{weldon_simple_1983}, similar to the optical theorem in zero-temperature field theory. In this picture, for the retarded self-energy, $\Im \Pi = -\omega(\Gamma^A - \Gamma^P)$ where $\Gamma^A$ and $\Gamma^P$ are the photon absorption and production rates resulting from processes that can be expressed with half of the loop diagram. The electrons and positrons from the loop are all put on shell, and are weighted by their phase space distributions (with or without a Pauli blocking factor depending on whether or not they are in the final state). The quantity $-\Im\Pi/\omega$ can thus be interpreted as the rate for the photon to equilibrate with the plasma. There are 8 processes contributing to the imaginary part, 4 absorption processes and 4 emission processes. However, the matrix elements are the same for each process and its reverse due to $CP$-invariance. The relevant processes are
\begin{itemize}
    \item Cherenkov absorption for positrons and electrons, which is allowed for spacelike photons with $n>1$,
    \item the decay of a photon to an electron-positron pair, which is only kinematically allowed for timelike photons with positive energy $\omega > 0$ and $\xi>1$,
    \item and absorption to the vacuum, which is only allowed for timelike photons with negative energy $\omega < 0$, and $\xi>1$.
\end{itemize}
The terms proportional to $-1$ from Eq.~\eqref{eq:minusone} do not diverge for the imaginary part. Furthermore, they are essential parts of the Pauli blocking factors involved. We obtain 
\begin{widetext}
\begin{equation}\label{ImPiE}
\begin{aligned}
    \Im \Pi =& -\frac{1}{16\pi k} \Bigg[ 
    \int_{E_{+-}}^{E_{++}} \dd{E_p} \qty|\mathcal{M}_{-}|^2 \qty(1-f_p-\bar{f}_p)\Theta(K^2-4m^2)\Theta(\omega)
    + \int_{E_{-+}}^{\infty} \dd{E_p} \qty|\mathcal{M}_{+}|^2 \qty(f_p+\bar{f}_p)\Theta(-K^2)\\
    +\ &\int_{E_{++}}^{\infty} \dd{E_p} \qty|\mathcal{M}_{-}|^2 \qty(f_p+\bar{f}_p)\Theta(-K^2)+\int_{E_{--}}^{E_{-+}} \dd{E_p} \qty|\mathcal{M}_{+}|^2 \qty(1-f_p-\bar{f}_p)\Theta(K^2-4m^2)\Theta(-\omega)
    \Bigg],
\end{aligned}
\end{equation}
\end{widetext}
where
\begin{equation}
    E_{\pm \pm} = \pm \frac{\omega}{2} \pm \frac{k}{2} \sqrt{1-\frac{4m^2}{K^2}} 
\end{equation}
and
\begin{align}
    \qty|\mathcal{M}^L_{\pm}|^2 &= 8 \pi \alpha \frac{K^2}{k^2}\qty(4E_p\omega \pm 4E_p^2 \pm K^2), \\
    \qty|\mathcal{M}^T_{\pm}|^2 &= 8 \pi \alpha \qty[\mp K^2 \mp2m^2- \frac{2K^2}{k^2}\qty(E_p\omega \pm E_p^2 \pm \frac{K^2}{4})].\nonumber
\end{align}
Restricting ourselves to $\omega>0$ (since the imaginary part of the self-energy is simply odd under $\omega \to -\omega$), the last term in Eq.~\eqref{ImPiE} vanishes. With $\gamma \equiv \frac{E_p}{m} = 1/{\sqrt{1-v^2}}$, we further define 
\begin{align}
    H_{\pm}^a &= \frac{\qty|\mathcal{M}_{\pm}|^2}{16 \pi \alpha m \omega n}\\ \nonumber \\ \qquad \gamma_\pm &= \abs{\frac{\omega}{2m}\qty[1 \pm \frac{k}{\omega} \sqrt{1-\frac{4m^2}{K^2}}]},
\end{align}
so that
\begin{align}
    \Im \Pi &= - \alpha m^2 \Bigg[ 
\int_{\gamma_{-}}^{\gamma_{+}} \dd{\gamma} H_{-} \qty[1-f_p-\bar{f}_p]\Theta(K^2-4m^2)\nonumber\\&+\int_{\gamma_{-}}^{\infty} \dd{\gamma} H_{+} \qty[f_p+\bar{f}_p]\Theta(-K^2)\nonumber\\&+\int_{\gamma_{+}}^{\infty} \dd{\gamma} H_{-} \qty[f_p+\bar{f}_p]\Theta(-K^2)
    \Bigg].\label{ImPigamma}
\end{align}
Once again we integrate by parts to obtain an analytic approximation. Defining
\begin{align}
    \Gamma_a(\gamma) &= \int_0^\gamma \dd{\gamma'} H^a_-(\gamma')\\
    &= 
    \begin{cases}
    \displaystyle \frac{\omega K^2}{k^3}\qty(\gamma^2-\frac{K^2}{2m\omega} \gamma - \frac{2m}{3\omega}\gamma^3) & a=L,\\
    \displaystyle \frac{m}{k} \qty(1+\frac{K^2}{2m^2})\gamma
    - \frac{K^2}{2} \Gamma_L(\gamma) & a=T,
    \end{cases} \nonumber
\end{align}
and realizing that $\int_0^\gamma \dd{\gamma'} H^a_+(\gamma')=\Gamma_a(-\gamma)$, we integrate by parts and apply the sharp-peak approximation, where the sharp peak occurs at $\gamma^* = 1/\sqrt{1-v^*}$. The boundary terms do not vanish at $\gamma = \gamma_\pm$, requiring that we evaluate the integral on a case-by-case basis in different kinematic regimes. 
    
For timelike photons, $n<1$, the only non-zero term in Eq.~\eqref{ImPigamma} is the first term. The Heaviside function has support only when $\xi > 1$ which corresponds to the regime where the decay $\gamma \to e^+ e^-$ is allowed. This is precisely the process that Ref.~\cite{braaten_neutrino_1993} avoided when dropping the $(K^2)^2$ term in the denominator (since this is process is non-physical for on-shell photons). Therefore, the imaginary part is exactly zero for light, timelike photons. For timelike heavy photons $(\xi > 1)$, integration by parts yields 
\begin{align}
   & \Im \Pi \simeq - \alpha m^2 \bigg[ \frac{2}{3}\qty(1+2\xi^2)\sqrt{1-\frac{1}{\xi^2}}+ \label{eq:ImPiTimelike}\\
    & \qty(f_+ + \bar{f}_+) \qty[\Gamma(\bar{\gamma}^*) - \Gamma(\gamma_+)]+ \qty(f_- + \bar{f}_-) \qty[\Gamma(\gamma_-) - \Gamma(\bar{\gamma}^*)] \bigg].\nonumber
\end{align}
where $\bar{\gamma}^* = \gamma_-$ if $\gamma^* < \gamma_-$, $\bar{\gamma}^* = \gamma_+$ if $\gamma^* > \gamma_+$, and $\bar{\gamma}^*={\gamma}^*$ otherwise. Notably, if the sharp peak at $\gamma^*$ falls outside of the limits of integration, the integrand will have the most support at the integration boundary that is closest to $\gamma^*$. For spacelike photons, the first integral from Eq.~\eqref{ImPigamma} vanishes and we obtain
\begin{align}\label{eq:ImPiSpacelike}
    \Im \Pi \simeq - \alpha m^2 \bigg\{&
    \qty(f_+ + \bar{f}_+) \qty[\Gamma(\gamma^*_+) - \Gamma(\gamma_+)]\\
    &+ \qty(f_- + \bar{f}_-) \qty[\Gamma(-\gamma^*_-) - \Gamma(-\gamma_-)]\bigg\}, \nonumber
\end{align}
where $\gamma^*_\pm = \gamma_\pm$ if $\gamma^* < \gamma_\pm$ and $\gamma^*_\pm = \gamma^*$ otherwise.

\subsection{Limiting behaviour}
\subsubsection{Kinematics}
In the low-$k$ limit ($n\rightarrow0$), the self-energy can be evaluated exactly for heavy timelike photons,
\begin{equation}
\begin{aligned}
    \Im \Pi_{L,T}& = - \alpha m^2 \frac{2}{3} (1+2\xi^2)\sqrt{1-\frac{1}{\xi^2}}\\
    &\times \qty(1 - f(E = \omega/2) - \bar{f}(E = \omega/2)).
\end{aligned}
\end{equation}
For spacelike photons, taking $n \to 1$ (or $\xi \rightarrow0$) sends $\gamma_+$ and $\gamma_-$ to infinity, so the integral for the imaginary part vanishes, $\Im \Pi_{L,T} (n \to 1) = 0$. This is expected from the fact that the limit $\xi \to 0$ corresponds to the on-shell self-energy, which has no imaginary part. Similarly, taking $n \to \infty$ (corresponding to $\omega = 0$) sets $\gamma_+=\gamma_-$ and $H_+ + H_- = 0$, so $\Im \Pi_{L,T} (n \to \infty) = 0$. The imaginary part of the self-energy also vanishes in the limit $\xi \to 1$ for similar reasons. 

\begin{figure*}[t]
    \centering
\includegraphics[width=0.8\textwidth]{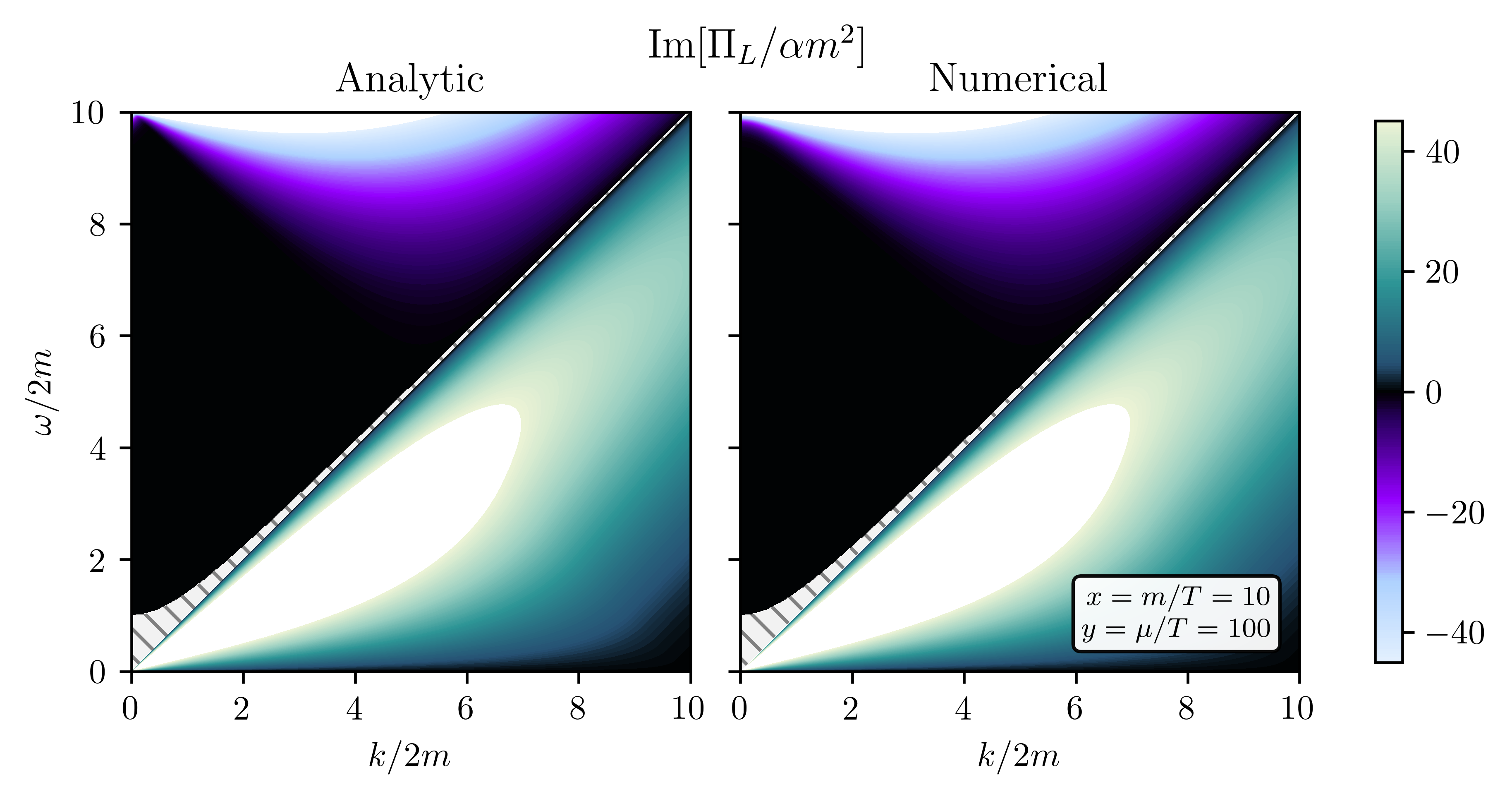}
\includegraphics[width=0.8\textwidth]{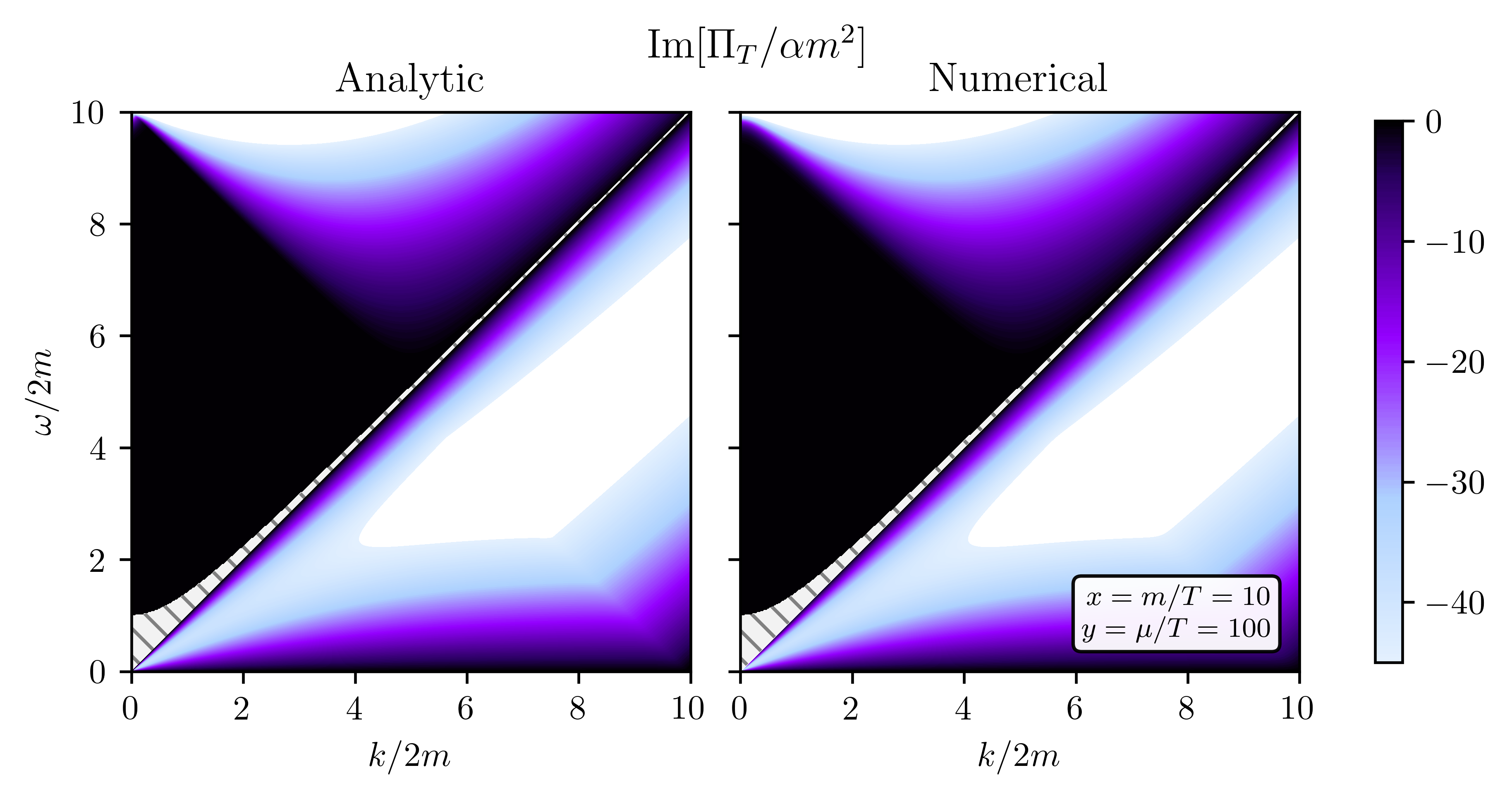}
\vspace{-0.5cm}
    \caption{The imaginary part of the self-energy as a function of $\omega$ and $k$, for longitudinal (top) and transverse (bottom) modes in a plasma where $x=10$ and $y=100$. The self-energy is computed with the analytic approximation (left panels) and the numerical integral of $H(v)$ (right panels). The dashed region indicates where the imaginary part is exactly zero, due to kinematics.}
    \label{fig:ImPi}
\end{figure*}

\subsubsection{Plasma Properties }
The imaginary part of $\Pi^{\mu \nu}$ depends on the plasma properties only through $v_*$, and considerations from the previous Section all apply. The expressions for the imaginary part are well-defined when taking $v_* = \sqrt{5T/m}$, or even $v_* \to 0$, in the classical case. They are also well-defined when taking $v_* = v_F$ in the high-$\mu$ case or $v_* = 1 - {m^2}/{8\sqrt{3}T^2}$ in the high-$T$ case. Taking $v_* \to 1$ (equivalently, $\gamma_* \to \infty$) in the high-$T$ case is well-defined for timelike photons, as can be seen from the definition of $\bar{\gamma}^*$. However, the spacelike expression diverges in the $v_* \to 1$ limit. One can instead express these in terms of finite $E_* = (m\gamma_*)$ and then take the limit $m \to 0$. 

Finally, our expression reduces to the Lindhard function in the degenerate limit with low Fermi velocity. Using the same procedure as in the real part, as well as $\alpha m^2 \simeq 3\pi \omega_p^2/4v_F^3$ at leading order in $v_F$, we find that our spacelike expression leads to:
\begin{equation}
    \Im[\epsilon_L] =
    \begin{cases}
        \frac{3 \pi \omega _p^2 \omega}{2 k^3 v_F^3}
        &\text{if }~ z+u<1 
        \\
        \frac{3 \pi \omega_p^2 p_F}{4 k^3 v_F^2} \qty(1-\qty(z-u)^2)
        &\text{if }~ \abs{z-u}<1<z+u
        \\
        0 &\text{if }  ~\abs{z-u}>1.
    \end{cases}
\end{equation}
The three cases above correspond to $\gamma_-<\gamma_+<\gamma_*$, $\gamma_-<\gamma_*<\gamma_+$, and $\gamma_*<\gamma_-<\gamma_+$, respectively. The expression once again matches exactly with the Lindhard formula~\cite{Lindhard_1954,Dressel_Gruner_2002}.

\subsection{Testing the approximation}
To compare our approximation with numerical integration, we show the imaginary part of the one-loop self-energy in Figure~\ref{fig:ImPi} for the same plasma parameters as Figure~\ref{fig:RePi}. The dashed regions indicate parts of phase space where the self-energy is exactly zero because none of the four processes involved is kinematically allowed. Crucially, the self-energy is non-zero far off shell when $\omega \gg k$ or for spacelike photons. 

The accuracy of the analytic approximation depends on the properties of the ambient plasma, just like in the case of the real part of $\Pi$. The approximation is again accurate at the sub-percent level for the plasma properties shown in Figure~\ref{fig:ImPi}, while the largest inaccuracies are seen in the high-$T$ regime $(x \ll 1)$ and can be of $\mathcal{O}(0.1-1)$ for timelike photons in regions where $\omega \sim k$ or for spacelike photons in regions where $k \gg \omega$. These discrepancies become especially important near regions where $\Im \Pi$ is close to zero, and they are once again largely driven by some shifts and distortions to the shape of the function which maintain the same overall qualitative structure. Therefore, we can expect that this approximation will also be most useful for performing phase space integrals, and will be less accurate for computing the self-energy at specific, local values of the energy and momentum.

\section{Discussion \& Conclusions}\label{sec:conclusion}
The photon self-energy is a necessary ingredient for performing calculations of interaction rates in the presence of a medium. The in-medium effects described by the self-energy can be useful for placing bounds on physics beyond the Standard Model in situations where the medium can enhance interaction rates do to e.g. resonant level-crossing between the photon and other states. The presence of a medium can even lead to qualitatively new channels of interaction, like the decay of photons into light, weakly coupled species. 

Braaten and Segel derived analytic approximations for the photon self-energy at one-loop, which are valid for soft photons that are close to being on shell. However, these approximations have frequently been used even when the photon is far from being on shell. Our results demonstrate that qualitatively new behaviour arises far off shell, for instance the real part of the self-energy is negative in some parts of phase space.

They key result of this work is the derivation of new analytic approximations for the photon self-energy, Eq.~\eqref{eq:RePi} for the real part and Eqs.~\eqref{eq:ImPiTimelike} and~\eqref{eq:ImPiSpacelike} for the imaginary part. These approximations are generally valid for both off-shell and on-shell photons, in all regions of phase space. These expressions are especially accurate in classical and degenerate plasmas, and can also reproduce the formula for the Lindhard dielectric function from solid-state physics (which also has applications to the search for sub-GeV dark matter, see e.g.~\cite{Knapen:2021bwg, Knapen:2021run, Hochberg:2021pkt}). For high-temperature plasmas, these expressions are less accurate when comparing at a particular point in phase space, but the differences are largely driven by small differences in the shape of features in areas where $\Pi$ is a steep function of $\omega$ or $k$. Therefore, we expect that integrals over phase space will still prove to be fairly accurate in the context of our high-$T$ approximation. Since the self-energy can be numerically slow to evaluate over all of phase space, we also expect that our approximations will be most useful for the purpose of performing such phase space integrals.

In some parts of phase space, the one-loop contribution to the self-energy is not sufficient to capture the relevant physics. For instance, in regions where $\Im[\Pi/\alpha m^2] \ll \alpha$ (or where it identically vanishes), higher-order contributions to the self-energy are known to dominate. Cutting these two-loop diagrams (analogous to the optical theorem) yields processes like Compton scattering and bremsstrahlung \cite{kapusta_finite-temperature_2011, redondo_solar_2013, An:2013yfc}. Our expressions do not capture these effects and therefore will not be a good approximation to the all-orders self-energy. 
Furthermore, in ultrarelativistic plasmas, higher-loop corrections for hard photons near the mass shell $(\omega^2-k^2 \sim \alpha T^2)$ are of the same order as the one-loop contribution because the photons can remain collinear with charged particles over multiple scattering times \cite{Landau:1953gr, Landau:1953um, Migdal:1956tc}. Again, in this regime our approximations are not expected to be representative of the all-orders self-energy.

In this work, we computed the self-energy for a homogeneous and isotropic plasma in thermal equilibrium, which is a good approximations to some astrophysical and cosmological environments as well as some solid-state materials. Our results can be generalized further in future work by dropping one or more of these assumptions, thus expanding the use of an accurate photon self-energy to an even broader range of environments.

\section*{Acknowledgements}
It is a pleasure to thank Nirmalya Brahma, Simon Caron-Huot, Charles Gale, Saniya Heeba, and Oscar Hern\'andez for useful conversations pertaining to this work and comments on the manuscript. We especially thank Tongyan Lin for pointing out the connection between our work and the Lindhard response function. HS was supported in part by a Master research scholarship from the Fonds de recherche du Qu\'ebec – Nature et technologies (FRQNT). KS and HS acknowledge support from the Programme \'Etablissement de la rel\`eve professorale from the FRQNT, from a Natural Sciences and Engineering Research Council of Canada Subatomic Physics Discovery Grant, and from the Canada Research Chairs program. KS thanks the Kavli Institute for the Physics and Mathematics of the Universe and the Kavli Institute for Theoretical Physics (supported by grant NSF PHY-2309135) for their hospitality in the late stages of the completion of this work. This analysis made use of \texttt{Numpy} \cite{harris2020array}, \texttt{Scipy} \cite{virtanen2020scipy}, \texttt{Matplotlib} \cite{hunter2007matplotlib}, and \texttt{Mathematica}~\cite{wolfram2003mathematica}.
\bibliography{main}
\end{document}